# Gigagauss magnetic fields generated via theta-pinching driven by multiple petawatt-class lasers

Huanyu Song[1,2], Zhengming Sheng[1,2,3*], Linzheng Wang[1,2], Min Chen[1,2], Suming Weng[1,2], Masakatsu Murakami[4], Jie Zhang[1,2,3,5]

[1] National Key Laboratory of Dark Matter Physics, Key Laboratory for Laser Plasmas (MOE), and School of Physics and Astronomy, Shanghai Jiao Tong University, Shanghai 200240, China

[2] Collaborative Innovation Center of IFSA, Shanghai Jiao Tong University, Shanghai 200240, China

[3] Tsung-Dao Lee Institute, Shanghai Jiao Tong University, Shanghai, 201210, China

[4] Institute of Laser Engineering, Osaka University, Osaka 565-0871, Japan

[5] Beijing National Laboratory for Condensed Matter Physics, Institute of Physics, Chinese Academy of Sciences, Beijing 100190, China

*Corresponding author: zmsheng@sjtu.edu.cn

**Abstract:** Extremely high axial magnetic fields above the gigagauss (GG) level are supposed to exist in neutron stars, which may be a one of the critical parameters for their internal structures and be responsible for the X and gamma-ray emission from these stars. Here we show that such ultrahigh magnetic fields can be produced by multiple petawatt-class lasers interacting with a cuboid solid target with a cylindrical microtube in the middle. It is found that the obliquely incident intense lasers at the target surfaces enable the produced hot electrons to form an azimuthal current and subsequently induce a seed magnetic field along the cylindrical axis inside the microtube as the hot electrons transport into it. This current-field configuration is similar to a theta-pinch device. When the hot electrons and energetic ions produced via target normal sheath acceleration converge towards the microtube axis, the seed magnetic field is dramatically amplified. This process continues until the magnetic pressure near the axis becomes comparable to the thermal pressure contributed both by hot electrons and energetic ions. Later on, as the plasma in the center start to be expelled outward by the magnetic pressure, an electron current ring with extremely high densities is formed, leading to a further boost of the magnetic fields to well above the GG-level. A scaling of the magnetic field strength with laser intensities, pulse durations, incident angles, and target sizes is presented and verified by numerical simulations, which demonstrates the robustness of our scheme. Our scheme is well suited for experimental realization on 100 terawatt-class to petawatt-class femtosecond or picosecond laser facilities with multiple linearly polarized laser beams.

# Introduction

Ultrahigh axial magnetic fields play an important role for a variety of high-energy-density phenomena in laboratory plasma physics[1–6] and astrophysics[7–10]. For example, in the fast ignition scheme of inertial confined fusion, tens of megagauss (MG) axial magnetic fields can greatly increase the coupling efficiency between the laser and the fusion core via magnetic collimation of relativistic laser-produced hot electrons[3]. When the axial magnetic field strength reaches the gigagauss (GG) level, a circularly polarized electromagnetic wave can propagate in extremely dense plasma in the whistler mode without encountering cutoff[4]. In a strongly magnetized plasma, an extreme case of the Faraday effect is reported, where a linearly polarized ultrashort laser pulse splits into two circularly polarized pulses of opposite handedness during its propagation[5]. Additionally, magnetic fields at the GG-class can trigger nonlinear QED effects under the interaction of a high energy electron beam[11,12]. In the context of astrophysics, magnetic field reconnection[7–9] in laboratory plasma with ultrahigh magnetic fields is also among current research interest. Extremely high axial magnetic fields at $10^{11} \sim 10^{13}$ gauss, typically exist in various neutron stars, such as pulsars and magnetars[13–16], can dramatically change the structure of neutral atoms and are often responsible for many astronomical observations[17–19]. If such high axial magnetic fields could be produced in laboratory, it would provide unique opportunities to study matters under such extreme conditions and thus improve our understandings of relevant astrophysical observations. However, the highest axial magnetic fields currently available in laboratories are at the MG-class and investigations in aforementioned areas have been largely limited to theoretical studies.

With the advent of high power lasers, intense laser-plasma interactions have opened up the possibility of creating high currents even exceeding the classical Alfven current limit[20], which may lead to the generation of ultrahigh magnetic fields at hundreds of MG class or even GG class[21–23]. However, electrons preferentially move along the direction of laser propagation, the generated magnetic fields are usually azimuthal and not suitable for applications that require axial fields. Therefore, various schemes for generating axial magnetic fields have been proposed, which include the use of the inverse Faraday effect[24–26], laser-driven coils[27–32], magnetic flux compression[33–35], and microtube implosion[36,37]. For example, through the inverse Faraday effect, GG-class axial magnetic fields could be produced from the interaction of extremely high-power circularly polarized (CP) lasers with overdense plasma[24] and intense spiral-profiled lasers with underdense plasma[26]. However, the generation of CP lasers from conventional linearly polarized (LP) lasers at high power requires a large size of waveplates and the produced magnetic fields are within a short duration. The use of specially shaped targets such as "escargot" target[28] or capacitor-coil target[31] is another route, but the highest magnetic fields experimentally observed to date are on the tens of MG-class[32]. However, the underlying mechanism of such a strong field is not fully understood[38]. In a recent study, Peebles *et al*. conducted extensive experiments to assess the potential of different types of laser-driven coils to generate strong magnetic fields at the MG-level[39]. Moreover, it is proposed to use several LP laser beams with their pointing directions twisted to generate tens of MG axial magnetic fields[40]. Amplifying an embedded seed magnetic field by magnetic flux compression is an alternative method to generate strong magnetic fields, since the magnetic flux is conserved, the amplified field is normally limited to about tens of MG. Up to now, the highest record of 70 MG was achieved via hydrodynamic implosion on the NIF facility[35]. Recently, a novel concept called the microtube implosion has been proposed[36], where an embedded 60-MG seed magnetic field can be amplified to the GG-class by collective Larmor gyromotions, enhancing its strength by 2 orders of magnitude. However, generating a high seed magnetic field at the level of a few tens of MG for that scheme is technically challenging.

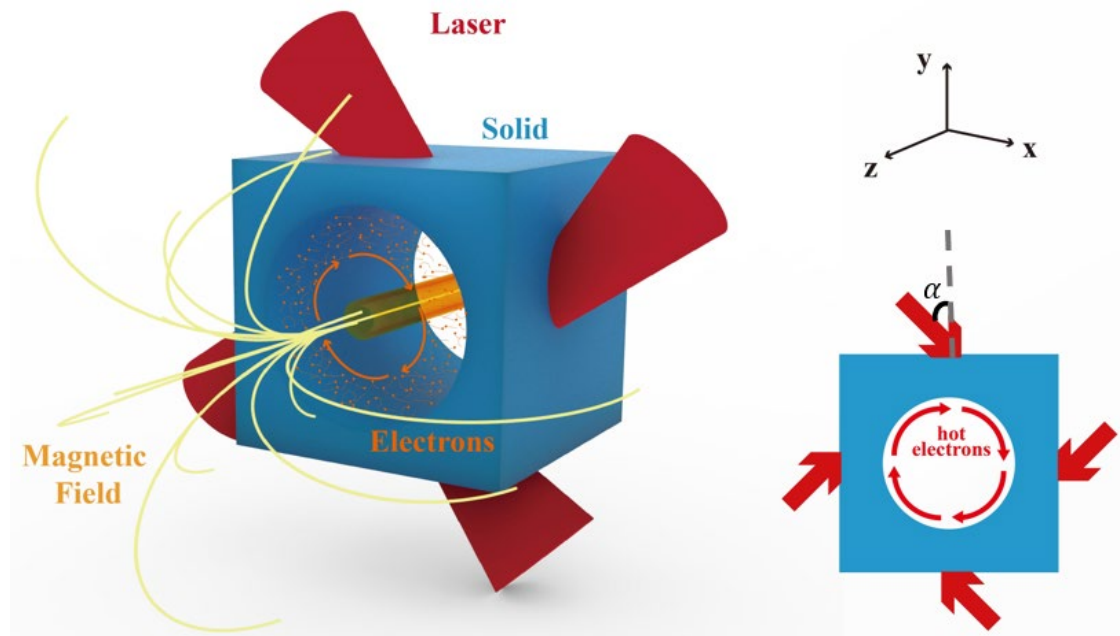

**Fig. 1** Perspective view of a microtube irradiated by four ultra-intense laser pulses with tilted pointing directions. Such a laser irradiation configuration leads the generation of an azimuthal current via the transfer of laser angular momenta to hot electrons, which produces an axial magnetic field. This forms a current-field configuration similar to a theta-pinch device. The insert diagram shows the cross-sectional in the 2D geometry, where red arrows illustrate the laser directions and azimuthal currents are formed by laser driven hot electrons inside the cylindrical hollow. The four LP laser pulses are incident respectively with the same angle $\alpha$ in four sides.

These studies mentioned above show that it is still technically challenging to generate GG-class quasi-static axial magnetic fields with existing schemes. In this work, we present an approach currently affordable with multiple hundreds of terawatt (TW) class or petawatt (PW) class lasers for the generation of GG-class magnetic fields. The interaction configuration is partially inspired by the cylindrical theta (θ) – pinch[41], where the key aspect of our design is to generate a current flow in the azimuthal direction and meanwhile squeeze inward the magnetic field via the hot electron transport. Our approach, as illustrated in Fig. 1, is realized by use of four intense laser pulses, which are incident obliquely onto the surfaces of a solid target from four sides simultaneously. The target is a cuboid with a cylindrical microtube in the middle. The generation of the high magnetic fields involves three stages. In stage I, relativistic hot electrons are produced near the target surfaces during the laser-plasma interaction. As they transport through the target into the cylindrical microtube, an azimuthal current is formed by overall angular momenta of the hot electrons due to the oblique incidence of lasers and subsequently a seed magnetic field pointing along the cylindrical axis is self-generated. In stage II, the seed magnetic field is quickly amplified and converges towards the tube axis as the hot electrons transport into the vacuum tube. In the meanwhile, ions from the tube wall are also accelerated via the target normal sheath acceleration (TNSA) mechanism[42]. The above processes continue until the magnetic pressure is balanced by thermal pressures, which corresponds to the plasma equilibrium in a theta-pinch device. Afterwards, as the hot plasma is expelled outward radially by the magnetic pressure, it is squeezed to form a ring with extremely high current density. Correspondingly, the magnetic field strength at the center is further boosted to the maximum value well above the GG-class. In stage III, the magnetic field starts to decrease due to the decay of the current ring and the appearance of diamagnetic currents.

## Results

**2D simulations.** To facilitate the demonstration of our scheme and reduce the computational cost, we have first performed two-dimensional (2D) particle-in-cell (PIC) simulations. Simulations have been performed using the massively parallel, fully relativistic, electro-magnetic PIC code EPOCH[43]. The full size of a computational box is 20 μm × 20 μm, which is sampled by 1500 × 1500 cells with 100 macro particles for aluminum ions and 200 macro particles for electrons. Here, we employ solid aluminum as the target material, with an initial ion density $n_{i0} = 40n_c$ and electron density $n_{e0} = 13n_{i0}$, where $n_c$ is the critical density corresponding to a laser wavelength $\lambda = 0.8$ μm. The ions are considered to be fully ionized, and physical electron-to-ion mass ratios are used. The cuboid target is placed at the center of the computational box with $D_0 = 14$ μm side lengths. The inner radius of the microtube is $R_0 = 5$ μm, so the minimum thickness is $\Delta R = 2$ μm. The four LP laser pulses are incident onto the target surfaces from four sides with an incident angle $\alpha = 0.4$ rad (or 23 degree). The temporal Gaussian envelope duration is $\tau = 50$ fs (FWHM) and ends at t = 100 fs, the peak intensity $I_L$ is $1.34 \times 10^{21}$ W/cm$^2$, corresponding to the normalized laser field amplitude $a_0 = 25$ for the laser wavelength $\lambda = 0.8$ μm. The laser peak power is about 2.6 PW. The laser pulses are co-timed such that the peaks of all four pulses interact with the target surfaces simultaneously.

The evolution of the plasma and the longitudinal magnetic field is illustrated in Fig. 2, which consists of three stages. In stage I, a seed magnetic field distributed within the microtube with certain spatial fluctuations is generated by the transfer of the angular momenta of lasers to the hot electrons (panels a – d in Fig. 2); In stage II, both hot electrons and ions are accelerated inward, leading to the convergence and amplification of the seed magnetic field at the center (panels e - h) until the magnetic pressure is balanced by the thermal pressure of hot electrons and ions. And then as the hot plasma is expelled outward by magnetic pressure later, a rotating hollow ring that rotates clockwise is formed due to the electrons carry angular momenta, which generates a compressed electron current layer in the azimuthal direction and subsequently a GG-class magnetic field (panels i - l). In stage III, the structure of hollow ring decays slowly and the magnetic field decreases with time (panels m - p) partially due to the formation diamagnetic currents, as discussed later. Snapshots of distributions of the normalized electron density $\tilde{n}_e = n_e/Zn_c$, the normalized ion density $\tilde{n}_i = n_i/n_c$, longitudinal magnetic field $B_z$ and the azimuthal current $J_\theta$ (at $y = 0$ µm) found in stages I, II and III are presented to capture the evolution processes. Here $n_c$ is the critical density for the corresponding laser wavelength.

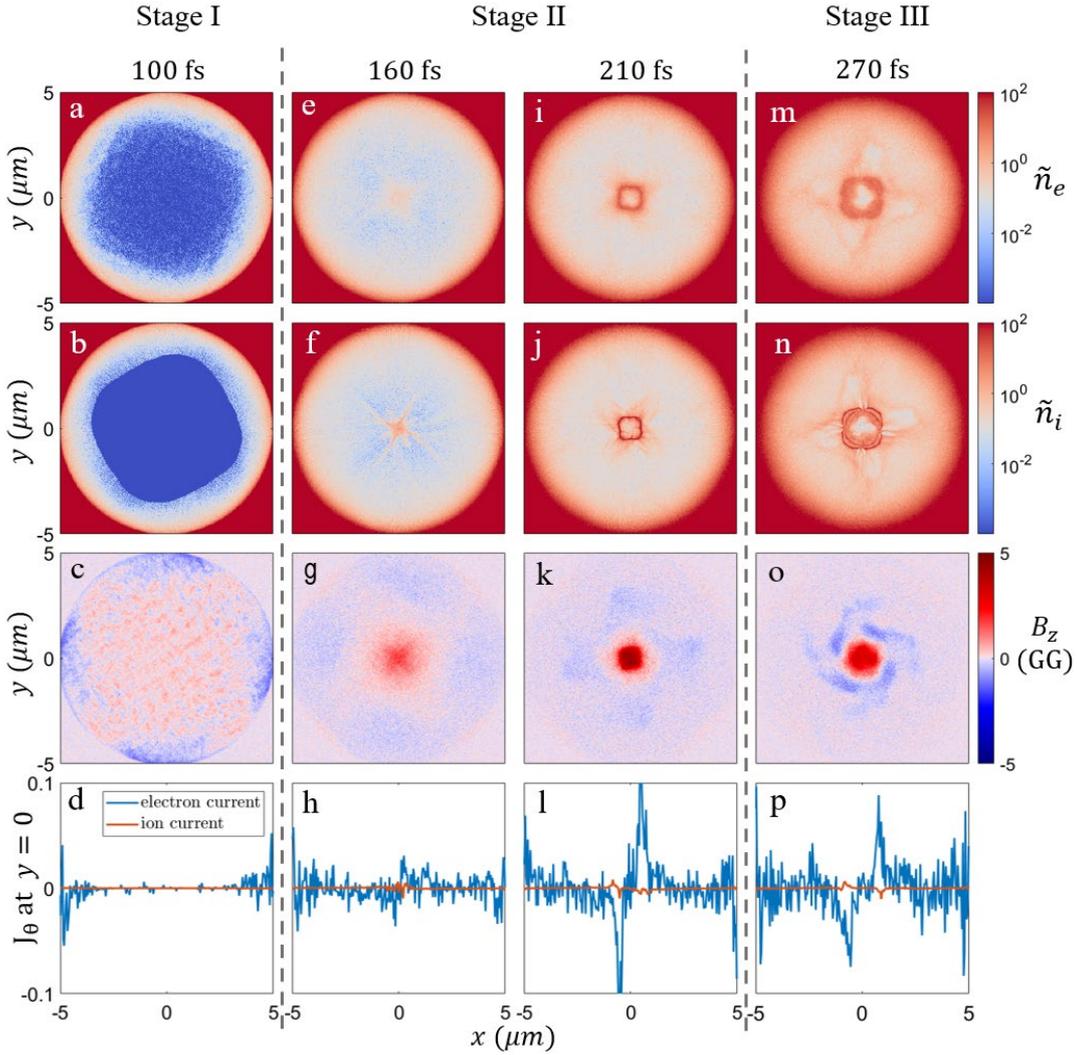

**Fig. 2** Simulation results at four different times, t = 100, 160, 210, and 270 fs (from left to right). **a, e, i,** and **m** The profiles of the electron density, where the density is normalized to the critical density $\tilde{n}_e = n_e/Zn_c$, where $n_c$ is the critical density. **b, f, j,** and **n** The profiles of the ion density, where the density is normalized to the critical density $\tilde{n}_i = n_i/n_c$. **c, g, k,** and **o** The profiles of the axial magnetic field, where the field is normalized to 1GG. **d, h, l,** and **p** The profiles of azimuthal current $J_\theta$ at $y = 0$ µm, the blue line is the electron current and the orange line is the ion current, where the current is normalized to $10^{15}$ A/cm$^2$.

In stage I, four obliquely incident laser pulses interact with the target and generate hot electrons. The specific spatial arrangement of laser pulses enables the transfer of a net angular momentum to the electrons and ions clockwise, which implies a rotating plasma environment. The angular directions of individual electrons are simply related with their kinetic energy and the experienced Coulomb potential changes[44]. Electrons with the same energy can move in different directions when they have experienced different Coulomb potentials during acceleration. For ions, due to significant mass disparity (where $m_i/m_e \approx 10^4$), they move at lower velocities and remain relatively localized as compared to electrons. Consequently, the rapidly rotating electrons quickly fill the microtube cavity (see Fig. 2a) and are responsible for the generation of the seed magnetic field (see Fig. 2c). As the ions rotate slowly and remain nearly localized (see Fig. 2b), they play a negligible role for the seed magnetic field. Correspondingly, the azimuthal ion current (at $y = 0$ μm) is nearly to be zero, as shown in Fig. 2d. Moreover, as hot electrons escape from the inner wall of the microtube in a clockwise direction, a high electrostatic potential is established quickly[30], dragging the background cold electrons towards the laser interaction volume in a counterclockwise direction. This cold electron current would create a reversed magnetic field at the edge of the microtube cavity (see the blue part in Fig. 2c).

It is worth mentioning that when lasers end at t = 100 fs, the self-generated seed magnetic field is about $B_{seed} = 1.08$ GG, which already exceeds the maximum magnetic field generated by other schemes[24,40]. In stage I, the generation of the seed magnetic field inside the cavity, the generation of the reversed magnetic field at the cavity edge due to the return current, and hot plasma implosion compression inside the cavity occur simultaneously and are coupled with each other. A quantitative analytical model for the seed magnetic field requires to consider the conservation of angular momentum[45,46], the feedback of hot electrons and plasma background electrons by implosion compression[47], which is currently not possible. We have performed additional simulation using normally incident lasers, and find that there is almost no seed magnetic field generated and subsequently no obvious magnetic field amplification as found in the following stages. We will discuss in detail the differences between normally incident and obliquely incident later.

In stage II, owing to the expansion of the inner wall plasma towards the center, the leading group of imploding electrons and ions reach the target center at t = 160 fs, but with a density two orders of magnitude smaller than the initial target density. Meanwhile, the self-generated seed magnetic field gradually converge to the central area within $r \sim 1$ μm, (see Fig. 2g), and its strength reaches 2.32 GG. Since the central plasma is dominated by this strong magnetic field, it is of interest to analyze the ratio between the thermal pressure $P_T = n_e k_B T_e + n_i k_B T_i$ and magnetic pressure $P_B = B^2/2\mu_0$. This is simply the beta factor given by

$$\beta = \frac{P_T}{P_B} = \frac{n_e k_B T_e + n_i k_B T_i}{B^2/2\mu_0}, \tag{1}$$

which is usually applied to illustrate the magnetic confinement of hot plasma. Especially, in the context of the theta-pinch, the beta factor is about 1 at the plasma equilibrium. In previous laser-based magnetic generation schemes[40,48], since the magnetic field is generally on the order of MG, the magnetic pressure is much smaller compared with thermal pressure ($\beta \gg 1$), and the role of magnetic pressure is negligible. In our scheme, however, the magnetic pressure matters when comparing the thermal pressure due to the GG-class magnetic field. To analyze $\beta$ more accurately, we take into account the electron temperature and ion temperature within $r < 1$ μm, which corresponds to the region of high magnetic fields.

Figures 3a and 3b show the electron energy spectrum and ion energy spectrum, correspondingly. The orange line corresponds to the full space energy spectrum, and the blue line corresponds to energy spectrum within $r < 1$ μm. One sees that the full space electrons differ from Maxwell–Jüttner (M–J) distribution for relativistic hot plasma[49], while the electrons within $r < 1$ μm follow M-J distribution, with a characteristic temperature about 7 MeV. For ions, the energetic ion spectrum has multiple energy components and the maximum energy reaches 1200 MeV, which is attributed to the fact that the ions are accelerated by the TNSA mechanism. In addition, one sees the high-energy ion tail is totally from the region of $r < 1$ μm. We can fit the ion distribution with two characteristic temperatures, i.e., $T_{i1} = 22$ MeV for the ions with energies below 500 MeV and temperature is $T_{i2} = 328$ MeV for the ions with energies above 500 MeV. On the other hand, the proportion of ions with energies below 500 MeV is about 0.78, and the proportion of ions with energies above 500 MeV is about 0.22. One can obtain the average ion temperature by $T_i = (0.78T_{i1} + 0.22T_{i2}) \approx$ 89 MeV, and the average electron density and ion density within $r < 1$ μm are 0.43 $\tilde{n}_e$ and 0.47 $\tilde{n}_i$, respectively. Consequently, at t = 160 fs, the total thermal pressure $P_T$ reaches $2.1 \times 10^7$ Gpa and the magnetic pressure $P_B$ is about $2.2 \times 10^7$ Gpa. Therefore, with a theta-pinch like process, the magnetic pressure is balanced by thermal pressures near the microtube axis with β = 0.95.

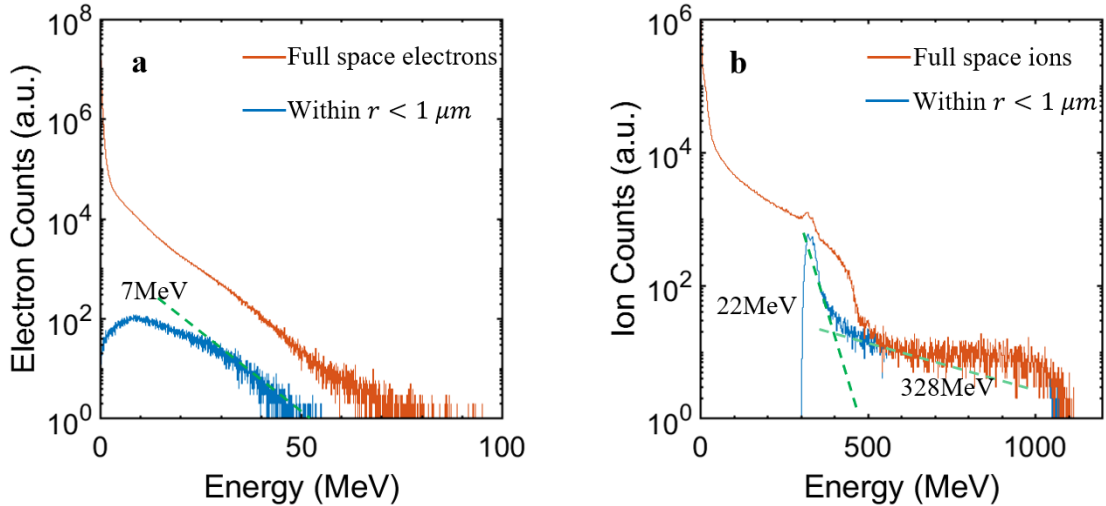

**Fig. 3** Electron energy spectrum (**a**) and ion energy spectrum at = 160 fs (**b**). The orange lines: the energy spectrum of full space electrons and full space ions. The blue lines: the energy spectrum for electrons and ions in the magnetic field region within $r < 1\,\mu m$. The green dotted lines: the characteristic temperature of electrons (7MeV) and ions (22MeV and 328MeV).

Since the magnetic pressure is so high that it is comparable to thermal pressure, the central electrons and ions are subject to radial outward force of the magnetic pressure. In a later stage, one can expect the central plasma will be expelled outward radially. Meanwhile, the cylindrically converging flow, composed of relativistic electrons and ions, is still imploding toward the center. As a result, at $t = 210$ fs, a hollow cylindrical structure of high plasma density is formed in the center. In the 2D geometry, the cross-section of the hollow cylindrical structure corresponds to a density hollow ring of radius $R_{ring} \sim 0.4$ μm, with practically no electrons and ions contained in the central area, as shown in Fig. 2j. By comparing the figures of $t = 160$ fs and $t = 210$ fs, one gains insight on how the hollow ring is formed as the central plasma expelled outward. It should be noted that the hollow rings of electrons and ions have the same radius, possibly due to the outward expulsion of the central plasma in a collective motion. Especially, the strong magnetic pressure is the key to generating the structure of the density hollow ring. It is found that if there is no seed magnetic field at the beginning with normally incident lasers, there is almost no obvious magnetic field generated at the center, and therefore no hollow ring forms. It is important to mention that our setup from the beginning produces a rotating plasma environment that rotates clockwise, the density hollow ring thus inherits the original angular momentum and rotation direction. Once the rotating hollow ring forms at $t = 210$ fs, a further boost of the magnetic field makes its strength at the center reach the maximum value about $B_{max} = 5.37$ GG (see Fig. 2k). Such a strong, well-defined magnetic field implies an equally well-defined current sustaining it. This is illustrated in Fig. 2l, where two current peaks are shown, corresponding to the azimuthal electron current (at $y = 0$ μm) with the current density reaching $10^{15}$ A/cm².

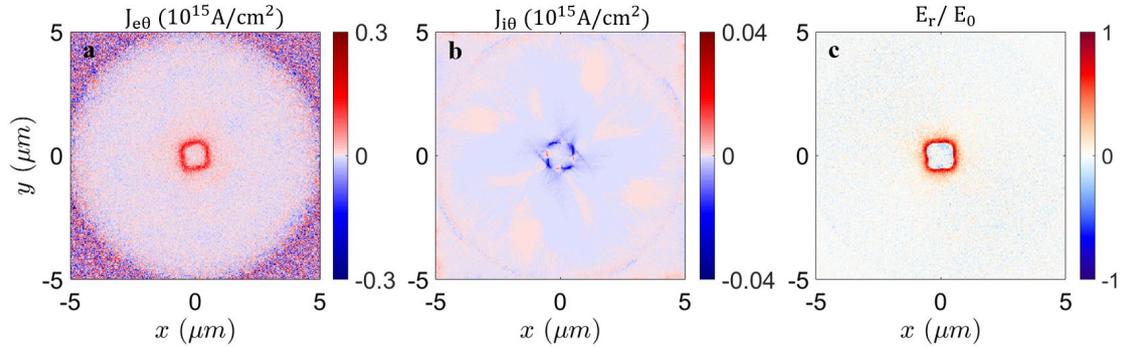

**Fig. 4 a** The azimuthal electron current density $J_{e\theta}$ at t = 210 fs. **b** The azimuthal ion current density $J_{i\theta}$ at t = 210 fs. **c** The normalized radial electric field generated by the no-quasi-neutral plasma at t = 210 fs, where $E_0 = 10^{14}$ V/m is the maximum laser electric field.

To further understand the azimuthal currents at $t = 210$ fs, we thus plot the whole electron current and ion current in azimuthal direction, as shown in Figs. 4a and 4b. Since both ions and electrons rotate clockwise, the resultant direction of the electron current $J_{e\theta}$ is opposite to that of the ion current $J_{i\theta}$. The electron current is on the order of $10^{15}$ A/cm², which is about 10 times of the ion current, suggesting that ions rotate slowly compared to electrons, and the role of ions is therefore negligible. Consequently, the rapidly rotating hollow ring produces strong azimuthal electron current which contributes dominantly to the generation of the ultrahigh magnetic field. It is worth mentioning that during the hot plasma implosion and magnetic field amplification processes, our scheme much differs from normal magnetic flux compression scheme using hollow cylindrical structures and pre-seeded magnetic fields. The magnetic flux in our scheme is not conserved, because the ultrahigh magnetic fields are generated by the azimuthal electron currents. On the other hand, the magnetic energy in our scheme increases rapidly because of the work done by the compressing forces.

It should be noted that the normalized densities of ions and electrons are not equal at each moment in Fig. 2. In particular, there is $\tilde{n}_i > \tilde{n}_e$ for the density near the center at $t = 210$ fs. From Gauss's law, there is an outward-pointing radial electric field with the non-quasi-neutral plasmas. Figure 4c shows the profile of radial electric field $E_r$ at $t = 210$ fs, its value is on the order of the maximum laser electric field. With the axial magnetic field $B_z$ and the radial electric field $E_r$, as well as the spatial gradient to $B_z$, the guiding center of the particle motion will perform $E \times B$ drift and grad-B drift motion, both along the azimuthal direction. Considering the $E_r$ and $B_z$ are both spatially and temporally inhomogeneous, the drift motion is more complicated in reality. However, due to plasma diamagnetism, any reactive magnetisation and drift currents will act to oppose the original magnetic field. Here, the diamagnetic drift motion tends to reduce the central magnetic field. Therefore, in stage III, the structure of density hollow ring gradually decays and becomes more complicated due to the drift motion, as shown in Fig. 2n. In fact, the central magnetic field begins to decrease coherently with the decay of the hollow ring. Meanwhile, the plasma in the environment of ultrahigh magnetic field can experience diamagnetic drift, resulting the reversed magnetic fields, as shown in Fig. 2o. When considering the profile of the azimuthal electron current at $y = 0$ μm (see Fig. 2p), we find that in addition to the two main peaks, there are also secondary peaks in opposite directions, which may correspond to the generation of the reversed magnetic fields. Additionally, simulation performed with different stages suggest the ion current is always much smaller than the electron current.

**3D simulations.** To further validate our scheme, we have carried out three-dimensional (3D) PIC simulation. The overall computational domain is x × y × z = 16 μm × 16 μm × 14 μm. The target and a single unit cell are both cubic with sizes of 14 μm × 14 μm × 14 μm and 25 nm × 25 nm × 25 nm, respectively. 6 particles for ions and 12 particles for electrons per cell. The other laser and plasma parameters are the same as those in above 2D simulations. Along the axial direction, periodic boundaries are used for both particles and fields. Thus, our simulations correspond to an infinitely long plasma microtube. Figures 5a and 5b show the perspective views of normalized electron density $\tilde{n}_e$ and the z-component of the magnetic field $B_z$ at $t = 210$ fs, respectively. To demonstrate clearly, the figures show a quarter of the full target volume. Comparing the 2D and 3D results indicates that the overall plasma behavior and the achieved key physical quantities, such as the maximum magnetic field and profile of electron density, agree with each other. For example, the hollow ring in Fig. 2i (2D simulations) corresponds to the hollow cylindrical structure in Fig. 5a (3D simulations), and the radius of both are approximately 0.4 μm. Meanwhile, 3D simulation shows that $B_{max}=$

5.95 GG, which is approximately equal to $B_{max}$= 5.37 GG found in the 2D simulations. Therefore, the physical mechanism demonstrated in the 2D simulations is reliable. It is found that the average magnetic field energy density exceeds $10^{16}$ J m$^{-3}$, the magnetic energy $\varepsilon_B$ within $r < 1$ µm is about 0.86 J $\left(\varepsilon_B = \int B_z^2/(2\mu_0)dV \approx 0.86 \text{ J}\right)$, and the total energy of the four beams is $\varepsilon_{laser} \approx 137$ J. The energy conversion efficiency from laser to the magnetic field is around 0.63%, which is 1~2 orders of magnitude higher than that for a laser-driven coil reported in Ref. [48].

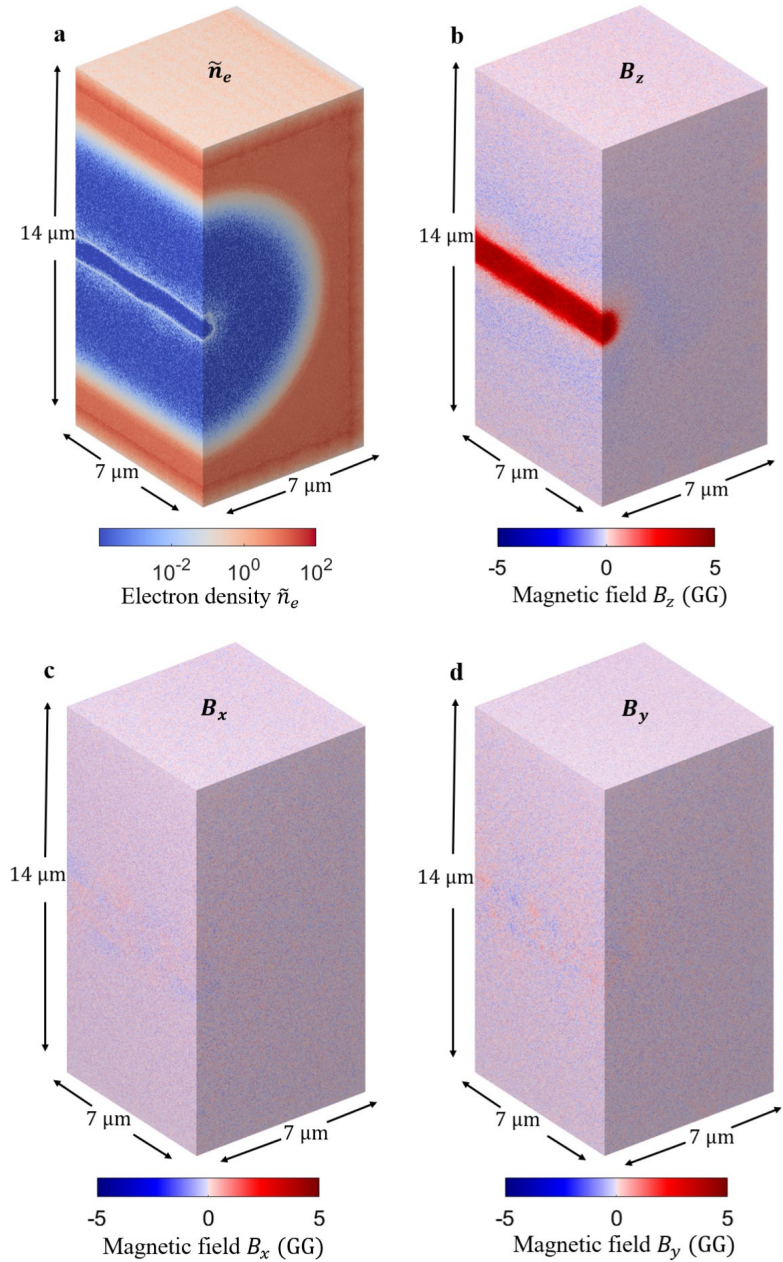

**Fig. 5** Perspective views of the normalized electron density $\tilde{n}_e$ (**a**), the z-component of the magnetic field $B_z$ (**b**), the x-component of the magnetic field $B_x$ (**c**) and the y-component of the magnetic field $B_y$ (**d**), respectively, observed at t = 210 fs obtained by a 3D simulation, where a quarter of the full target volume is shown.

The x-component of the magnetic field $B_x$ and the y-component of the magnetic field $B_y$ are shown in Fig. 5c and 5d respectively. It can be clearly seen that there are almost no $B_x$ and $B_y$ components. Therefore, our scheme can generate an almost pure axial magnetic field without angular magnetic field. In order to further compare the results of 2D simulations and 3D simulations, we have performed a series of 3D simulations by varying the normalized laser field amplitude $a_0$, while other parameters keep the same. The results given in Fig. 6 show that, the 2D and 3D performances of magnetic field generation in different $a_0$ are similar, for example, with the increase of $a_0$, the maximum magnetic field strength increases in both 2D and 3D simulations. Hence, our scheme indeed works under the 3D configuration with practical target and different laser conditions. However, owing to limitations in our computational ability, the cell size and particle numbers assigned to a cell in 3D simulations are substantially coarser than those treated in 2D simulations, which result in the magnetic field intensities of the 3D simulations being slightly higher than that of the 2D simulations. The dotted line in Fig .6 represents the theoretical results, which will be discussed in detail in the following section.

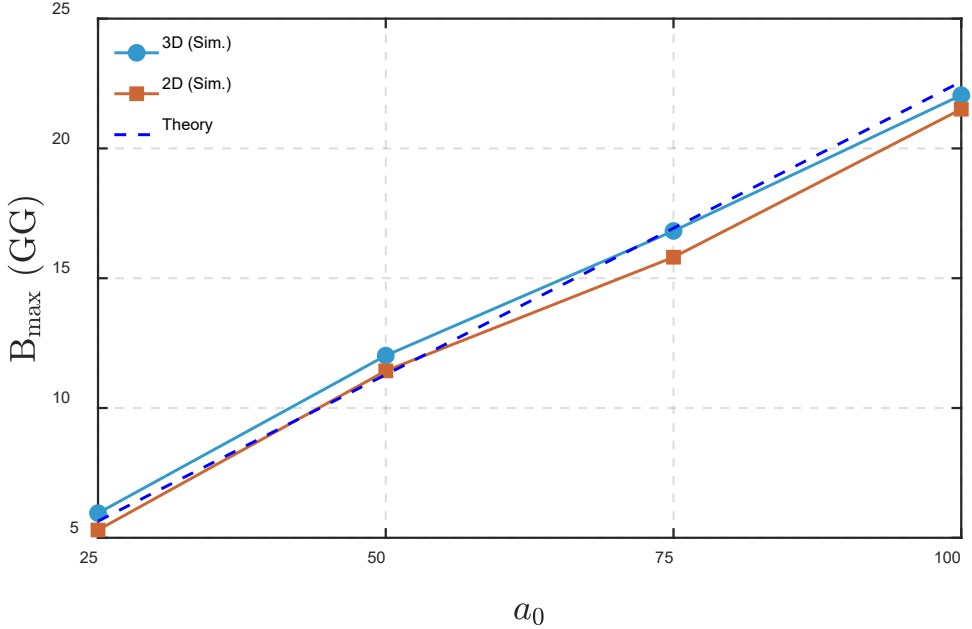

**Fig. 6** The maximum magnetic field as a function of the peak laser field strength, where the result predicted by the theory model given in Eq. (5) is compared with numerical simulation results obtained by2D and 3D simulations.

**Theoretical model.** To construct a theoretical model for our scheme, we begin by estimating the azimuthal electron current $J_{e\theta}$, which is due to $B_{max} \propto J_{e\theta} R$. At the time of the maximum magnetic field, the electrons that form the azimuthal current carry the total angular momentum as $L_e = N_e[\mathbf{r} \times \mathbf{p}] \sim R\gamma_a m_e J_{e\theta}/e$, where $N_e$ is the number of hot electrons contributing to the azimuthal current. We thus obtain that

$$J_{e\theta} = \frac{eL_e}{\gamma_a m_e R} \qquad (2)$$

where $\gamma_a = \sqrt{1 + a_0^2/2}$ is the relativistic gamma factor associated with the normalized laser amplitude $a_0$ for a linearly polarized laser pulse. Equation (2) tells us that one can obtain $J_{e\theta}$ and $B_{max}$ when one knows $L_e$. In Ref. [40] where the magnetic field is totally determined by the angular momenta transfer from lasers to electrons, the assessment of $L_e$ is through the application of angular momentum conservation principles. Once the number of absorbed photons is known, the angular momentum transferred to electrons can be obtained. However, the physical processes in our scheme are more complicated, the angular momenta transfer from lasers to electrons just serve to initiate the seed magnetic fields, not to determine the final magnetic field. Afterwards, the angular momenta of electrons at the time of maximum magnetic field are determined by the total electron flux emitted from the inner surface of the microtube[36], i.e., a phenomenological expression as

$$L_e \propto N_e \, \varepsilon_e, \tag{3}$$

where $\varepsilon_e$ is averaged energy of hot electrons. In our scheme, the total laser energy along the normal direction can be written as $U_L \sim 4I_L \tau_L D_0^2 \cos\alpha$ ($\alpha > 0$), assuming that the absorbed laser energy is uniformly transferred to the electrons contained in the target as[37,50]

$$\varepsilon_e = \frac{4\eta_a I_L \tau_L D_0 \cos\alpha}{(D_0^2 - \pi R_0^2)N_e}, \tag{4}$$

where $\eta_a$ and $D_0$ denote the laser absorption efficiency and the length of each side of the cuboid target, respectively. Since $B_{max} \propto J_{e\theta} R$, from Eqs. (2) – (4), the scaling for the maximum magnetic fields is obtained in terms of the laser and target parameters as

$$B_{max}(GG) = 0.053 \frac{\eta_a a_0 \tau_L D_0 \cos\alpha}{(D_0^2 - \pi R_0^2)\lambda_L^2}, \tag{5}$$

where $D_0$, $R_0$ and $\lambda_L$ are normalized to 1 μm, $\tau_L$ is normalized to 1 fs, and the numerical coefficient 0.053 is a fitting constant to the simulations. In principle, $\eta_a$ is bounded as $0.1 < \eta_a < 0.8$, and the absorption mechanism is more complex[51–54]. To simplify the discussion, our model ignores the dependence of $\eta_a$ on such parameters like $a_0$ and $\alpha$, and we find $\eta_a = 0.5$ agrees well with our simulation results with fitting constant 0.053. In accordance with Eq. (5), the manipulation of the axial magnetic field is achievable by altering laser intensity $a_0$, the duration of laser $\tau_L$, laser incident angle $\alpha$, and microtube radius $R_0$. Furthermore, it is found that the magnetic field decreases as the incident angle increases. This can be explained as follows. Although a larger incident angle may provide a higher angular momentum, the laser energy along the normal direction will decrease, resulting in a reduced compression effect of the magnetic field. On the other hand, when the incident angle is 0, there is no angular momentum. Therefore, Eq. (5) is applicable when $\alpha > 0$ rad.

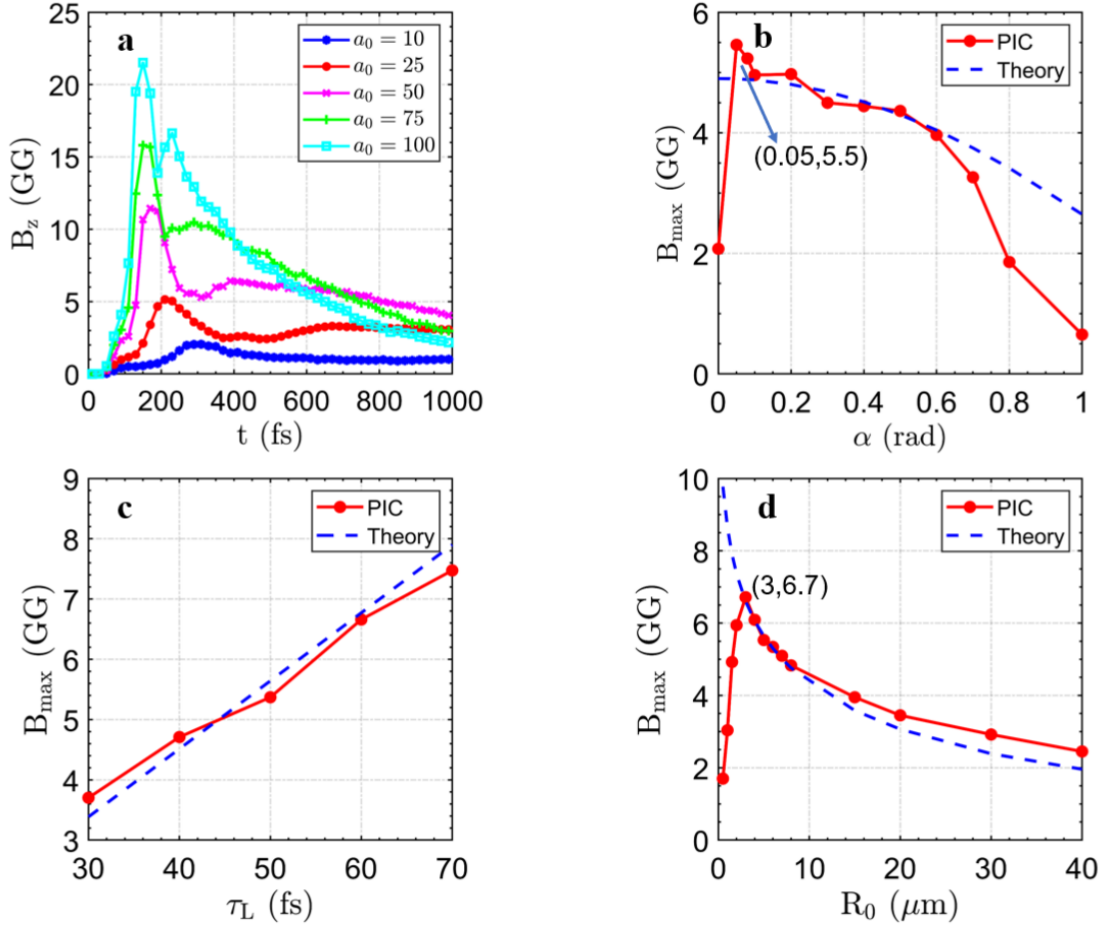

**Fig. 7 a** Temporal evolution of the central magnetic field, obtained from 2D simulations, under different laser intensities $a_0$, and other parameters are $R_0 = 5$ μm, $D_0 = 14$ μm, $\alpha = 0.4$ rad, $\tau_L = 50$ fs, the inset diagram show the fits based on Eq. (5). **b** Theoretical model (blue dashed line) for varying $\alpha$ and the maximum magnetic field from the simulations results (red circles), other parameters are $R_0 = 5$ μm, $D_0 = 14$ μm, $a_0 = 20$, $\tau_L = 50$ fs. **c** Theoretical model (blue dashed line) for varying $\tau_L$ and the maximum magnetic field from the simulations results (red circles), other parameters are $R_0 = 5$ μm, $D_0 = 14$ μm, $a_0 = 25$, $\alpha = 0.4$ rad. **d** Theoretical model (blue dashed line) for varying $R_0$ and the maximum magnetic field from the simulations results (red circles), other parameters are $D_0 = 2 \times (R_0 + 2)$ μm, $a_0 = 25$, $\alpha = 0.4$ rad, $\tau_L = 50$ fs.

To investigate the robustness of this model, we have performed a series of 2D simulations by varying the four parameters $a_0$, $\alpha$, $\tau_L$, and $R_0$, respectively. Figure 7a shows the temporal evolution of the magnetic field generation with different $a_0$, and other parameters are $R_0 = 5$ μm, $D_0 = 14$ μm, $\alpha = 0.4$ rad, $\tau_L = 50$ fs. It is found that the maximum magnetic fields strongly depend on the applied laser intensity, and the duration of magnetic fields is approximately 100 fs (FWHM), much longer than the laser pulse duration (~50fs). Most importantly, although the magnetic field decreases after reaching the maximum, the magnetic field strength is still above the GG level at $t = 1$ps, the lifetime of the magnetic field is one order of magnitude longer than the laser duration. Furthermore, as the laser intensity increases, the magnetic field is generated earlier. This is because hot electrons with a higher temperature can fill the microtube cavity quickly, resulting the generation of density hollow ring in advance. Even at lower laser intensity, e.g., $2.15 \times 10^{20}$ W/cm² ($a_0 = 10$), the maximum magnetic field strength is about 2-GG level, which is 1~2 order of magnitude higher than other schemes with same laser intensity[24,40,48].

Within the current laser facility capabilities, it is of particular interest to analyze the performance of our scheme in hundreds of TW-class. Figure 8a shows the simulation results for $I_L = 5.50 \times 10^{19}$ W/cm² ($a_0 = 5$), where the corresponding laser peak power is about 100 TW and total energy is about 10 J. It is found that the maximum magnetic field strength can still reach 1-GG level, which is consistent with our theoretical model (1.08 GG). In addition, the lifetime of the magnetic field far exceeds 1ps. The inset diagram in Fig. 8a shows that in a wide range of $5.50 \times 10^{19} \leq I_L[W/cm^2] \leq 2.15 \times 10^{22}$ ($5 \leq a_0 \leq 100$), the 2D simulation results are well approximated by the scaling relation in Eq. (5).

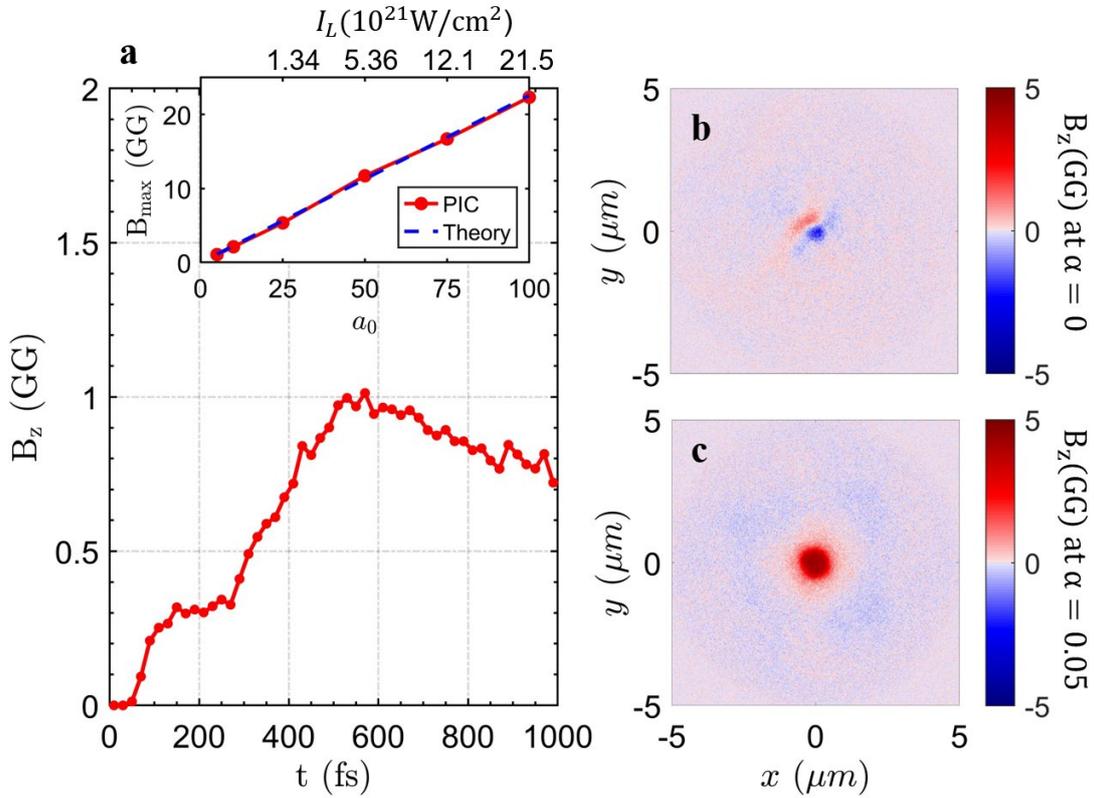

**Fig. 8 a** Temporal evolution of the central magnetic field under $a_0 = 5$, the corresponding peak power is 100 TW, the inset diagram shows the fits based on Eq. (5). **b** When $\alpha = 0$ (normal incidence), the profiles of the axial magnetic field $B_z$ at $t = 250$ fs, this case corresponds to the first point (0 rad, 2.0 GG) in Fig. 7b. **c** When $\alpha = 0.05$ rad (or 2.86 degree), the profiles of the axial magnetic field $B_z$ at $t = 200$ fs, this case corresponds to the maximum point (0.05 rad, 5.5 GG) in Fig. 7b.

The incident angle scan, shown in Fig. 7b is performed for a fixed laser intensity of ($a_0 = 20$). Note that within the range of $0.1$ rad $< \alpha < 0.7$ rad, the magnetic field strength varies inversely with incident angle, which is in good agreement with the simulation results. A small incident angle is simply to initiate the self-generated seed magnetic field, so that more laser energy can be used to increase the magnetic field during the implosion compression process. When $\alpha > 0.7$ rad, there has been a significant decrease in the magnetic field strength compared to the theoretical model, which means too large incident angle is not suitable for the generation of magnetic field. For $\alpha \leq 0.1$ rad, we scan five sets of $\alpha$, that is $\alpha = 0, 0.05, 0.08,$ and $0.1$ rad. It is found the simulation results slightly higher than our theoretical model, and $\alpha = 0.05$ rad (or 2.9 degree) is the most optimized incident angle, the magnetic field can reach 5.5 GG and the profile of magnetic field is shown in Fig. 8c. Moreover, as shown in the first point of Fig. 7b, one can find in the case of normal incidence, that is $\alpha = 0$, the magnetic field intensity is not strictly equal to 0. Since each of the inward hot electron flows carries its own internal magnetic field, these fields would interact and superimpose even at a normal laser incidence, leading to an unexpected magnetic field, as shown in Fig. 8b. Even though the magnetic field strength is about 2 GG, the distribution of the magnetic field is disordered, both positive polarity ($B_{max} > 0$) and negative polarity ($B_{max} < 0$) of magnetic fields occur simultaneously.

The dependence of the maximum magnetic field on pulse duration $\tau_L$, shown in Fig. 7c with red circle markers, matches well with the dependence given by Eq. (5) and shown with the blue dashed line. The laser pulse duration is believed to affect the number of hot electrons and, as a result, the magnetic field generation. In addition, the number of hot electrons is also strongly dependent on the pre-plasma density scale length[52,55,56], a pre-plasma is expected to lead a similar scaling multiplied by higher numerical factors. To avoid excessive expansion of our current paper, we leave the details for future studies. Figure 7d shows the effect of varying $R_0$. Due to the limitation of computing resources, the maximum microtube radius is up to $R_0 = 40$ μm. Within the range of $R_0 > 3$ μm, it indicates that a small radius is capable of supporting high magnetic fields, consistent with our model given in Eq. (5). Even at $R_0 = 40$ μm, the magnetic field can still beyond 2 GG. It should be noted that in cases of large radii, the simulation results are slightly higher than the theoretical model. This might be because we assumed a fixed absorption coefficient in the theoretical model. On the other hand, for the range of $R_0 < 3$ μm, the PIC simulation results deviate from the theoretical model. This may be due to the fact that the acceleration distance is too short to compress the magnetic field sufficiently. Therefore, $\alpha = 0.05$ rad and $R_0 = 3$ μm are the most optimized choice with our simulation parameters.

**Discussion.** We here briefly discuss laser requirements for our scheme. The laser duration in our calculations is within 30 ~ 70 fs and the target size is within 10 ~ 20 μm. To achieve GG-class magnetic fields experimentally, a rough estimate assuming a pulse duration of ~ 50 fs suggests that a laser system with total pulse energy of 10 J~ 100 J and a total power of 100 TW~ PW is required. Currently, femtosecond multiple PW-class laser systems like ELI-Beamlines[57] and ELI-NP[58] may provide necessary conditions for experimental demonstration. In addition, laser facilities like LFEX[59], NIF ARC[60], and Shengguang II PW systems can afford multiple PW laser beams on a picosecond timescale. Such long picosecond duration could match well with a larger target size, e.g., a few hundred microns, resulting in higher magnetic fields that occupy a larger volume. Therefore, it is expected to achieve an even higher magnetic field strength, e.g., sub-Tera-Gauss (TG), by proper designs of the experiments by matching the laser intensity, pulse duration, laser spot size, and incident angle with the target parameters (including the dimension of the target surface and inner microtube radius). Particularly, if the drive laser power reduces to the 100TW class, one can still obtain GG-class magnetic fields as the field strength scales linearly with the square-root of the laser intensity. Thus, with the existing PW-class laser systems, if a single beam is split into multiple ~100 TW beamlets[61], they are enough to be applied to test our scheme experimentally.

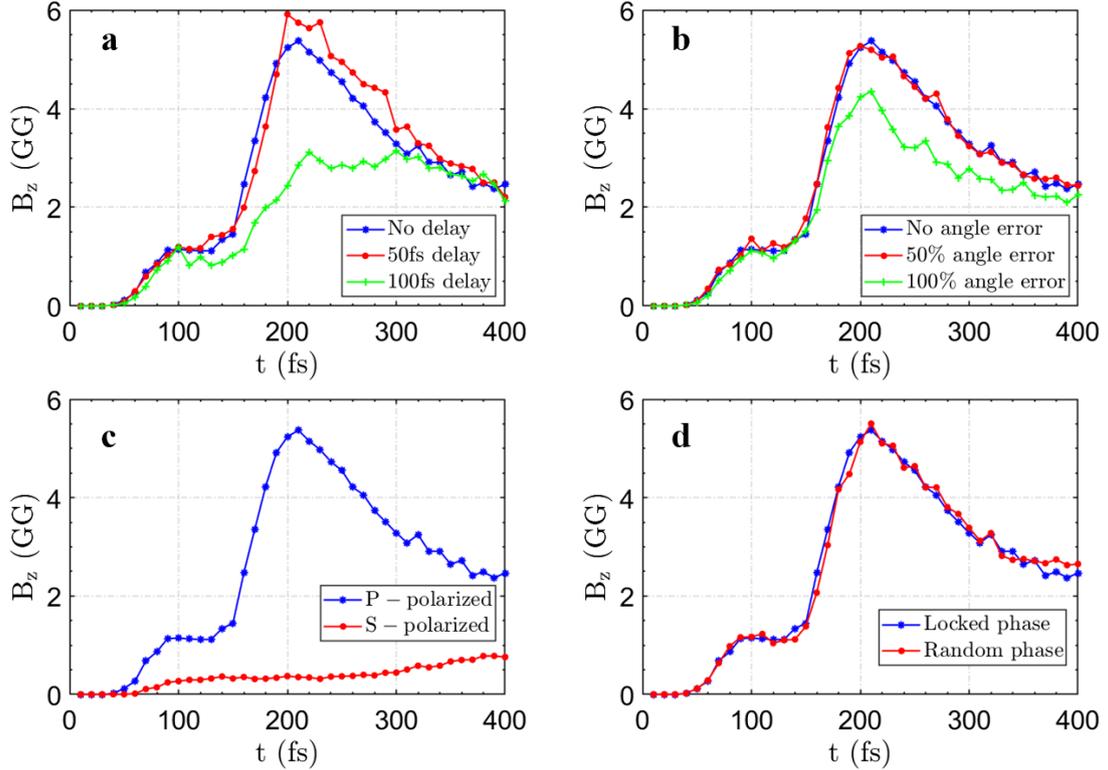

**Fig. 9** Assessment on the impact of time delay, incidence angle error, polarization direction and phase when $a_0 = 25$. **a** Time delay scenario considered in the simulations, the duration of lasers is 50 fs. 50 fs (100 fs) delay: one of the lasers has 50 fs (100 fs) time delay compared with other three lasers, respectively. **b** Incidence angle error scenarios considered in the simulations, the incidence angle is 0.4 rad, 50% (100%) angle error means the incidence angle of one of the lasers is 0.6 (0.8). **c** The cases of p-polarized lasers and s-polarized lasers are given by the blue and red solid lines, respectively. **d** Different phase scenarios considered in the simulations. Locked phase: the phases of all four lasers are set to 0. Random phase: the phases of all four lasers are random. These four additional simulations were performed while keeping all other parameters the same as those in Fig. 2.

Note that we have so far assumed that the lasers are perfectly co-timed. In experiments, however, laser synchronization is difficult in multi-laser facilities. Therefore, Assessing the impact of time delay is essential. Moreover, it is valuable to examine various aspects of our scheme, such as incidence angle error, polarization direction, and phase, that are likely to be important during experimental implementation at multiple laser facilities. Figures 9a and 9b provide the temporal evolution of the axial magnetic field for the time delay and incidence angle error, respectively. Short time delay, such as 50 fs delay or incidence angle error less than 50%, has almost no effect on the generation of the axial magnetic field, this provides encouraging prospects for experimental implementation. Even with a long delay (100 fs) or large angle error (100%), the magnetic field still reaches the 3-GG level, demonstrating that our scheme remains effective under sufficiently long delay conditions and large incidence angle error. In particular, four p-polarized beams are necessary for generating the axial magnetic field in our scheme (see Fig. 9c), this is because the production of hot electrons is directly related to p-polarized beams. Figure 9d shows that the introduction of random phases does not cause any significant changes, no phase control is required for the combination of multiple laser pulses in our scheme. Considering the realistic laser prepulse in experiment, we also evaluate the ionization effect for the generation of magnetic field when $a_0 = 25$. It is found that the maximum magnetic fields are insensitive to ionization states when the initial ionization states exceed $Al^{+5}$.

Detecting GG-order magnetic fields inside plasma poses challenges for traditional techniques that rely on charged particle sources. Laser-plasma interactions are accompanied by the generation of ultrahigh magnetic fields, thus ongoing efforts have been made to develop alternative techniques for inferring the existence of strong B-fields within dense plasmas, for example, an XFEL photon beam with Faraday rotation effect[62] and spin-polarized neutrons[63]. In addition, it is worth mentioning the practical application of ultrahigh magnetic fields. When the strength reaches $B \gg m_e^2 e^3 c/\hbar^3 = 2.35$ GG, the Coulomb force on an electron acts as a small perturbation compared to the magnetic force, the electron cyclotron energy $\hbar\omega_c$ becomes comparable to the atomic binding energy (the Rydberg), and thus the properties of matter are drastically modified by ultrahigh magnetic fields[64,65]. These phenomena mentioned above are currently under active investigations in astrophysics due to the extreme-field environment is typically found on the surfaces of neutron stars. In particular, our scheme sets a possible platform for reproducing extreme magnetic environment in laboratories using readily available laser facilities. In addition, our scheme can work with different numbers of laser beams, such as three beams or five beams. In this case, the outline of the target should be changed to triangle or pentagon in order to match the beam numbers. The key physics remains the same as described above.

In summary, we have demonstrated a novel mechanism to generate axial GG-level magnetic fields using multiple linearly polarized petawatt-class lasers. By tilting the laser irradiation directions, self-generated seed magnetic fields are filled in the whole vacuum tube of the target when laser-produced hot electrons propagate into it. Later, both hot electrons and ions accelerated via TNSA converge towards the target center, which is followed by the convergence and amplification of seed magnetic fields until the magnetic pressure becomes comparable to the thermal pressure. This forms a theta-pinch like equilibrium. Afterwards, the central hot plasma is expelled outward, which is accompanied with the formation of a plasma density hollow ring with a strong azimuthal current to maintain the ultrahigh magnetic fields before decay. At this stage, the peak magnetic fields can be well above a few GG level, where the magnetic field energy density is greater than $10^{16}$ J m$^{-3}$. Even at the 100TW-class lasers, the magnetic fields can still reach the GG level. The PIC simulations and supporting theory indicate our mechanism is robust and can be realized for a wide range of laser intensities, incident angles, laser durations and target parameters. Our scheme may provide a feasible way to produce ultrahigh magnetic fields even well beyond the GG-class for various applications.

## Acknowledgements

This work was supported by the Strategic Priority Research Program of Chinese Academy of Sciences (Grant Nos. XDA25050100 and XDA25010100) and the National Natural Science Foundation of China (Grant Nos. 12135009 and 11991074). PIC simulations were performed using EPOCH, developed under UK EPSRC (Grant Nos. EP/G054940, EP/G055165, and EP/G056803).